\documentclass[aps,prb,twocolumn,groupeauthor,showpacs]{revtex4}

\usepackage{amsmath}
\usepackage{graphicx}
\usepackage{dcolumn}
\usepackage{bm}
\usepackage[dvips]{color}
\usepackage{ulem}
\usepackage{multirow}

\begin{document}

\title{Exchange scattering as the driving force for ultrafast all-optical\\ and bias-controlled reversal in ferrimagnetic metallic structures}

\author{A. M. Kalashnikova and V. I. Kozub}\email{ven.kozub@mail.ioffe.ru}
\affiliation{Ioffe Physical-Technical Institute of the Russian Academy of Sciences, 194021 St. Petersburg, Russia}

\date{today}

  \begin{abstract}
  Experimentally observed ultrafast all-optical magnetization reversal in ferrimagnetic metals and heterostructures based on
  antiferromagnetically coupled ferromagnetic $d-$ and $f-$metallic layers relies on intricate energy and angular momentum flow between electrons, phonons and spins. Here we treat the problem of angular momentum transfer in the course of ultrafast laser-induced dynamics in a ferrimagnetic metallic system using microscopical approach based on the system of rate equations. We show that the magnetization reversal is supported by a coupling of $d-$ and $f-$ subsystems to delocalized $s-$ or $p-$ electrons. The latter can transfer spin between the two subsystems in an incoherent way owing to the $(s;p)-(d;f)$ exchange scattering. Since the effect of the external excitation in this process is reduced to the transient heating of the mobile electron
  subsystem, we also discuss possibility to trigger the magnetization reversal by applying a voltage bias pulse to antiferromagnetically coupled metallic ferromagnetic layers embedded in point contact or tunneling structures. We argue that such devices allow controlling reversal with high accuracy. We also suggest to use the anomalous Hall effect to register the reversal, thus playing a role of reading probes.

\end{abstract}


\maketitle

\section{Introduction}
The possibility of ultrafast control of the magnetic state of
nanostructures is an important constituent of ferromagnet-based
spintronics. Due to the problem of non-locality and difficulty to create strong yet short magnetic field pulse,\cite{Back-Science1999,Gerrits-Nature2002,Tudosa-Nature2004} the natural idea to use the latter is becoming incompatible with the request for further increase of storage densities and operation speed of novel spintronic devices.\cite{Stoer-book}
Thus a great enthusiasm  arose following the proposal \cite{Slon,Berger} to use spin injection for controlling the magnetization state of ferromagnetic
specimen with the help of an applied voltage. Later such a possibility
was profoundly studied both theoretically and experimentally (for
the critical review see e.g. [\onlinecite{me1}]). However, two
important obstacles were found. First, the switching time of the
magnetization reversal by spin injection is defined by
magnetization precession damping and is rather long
(around $\sim 10^{-9}$ s). Second, relatively large currents required
for effective switching inevitably lead to unreasonable heat
losses.

Thus a great attention\cite{Kirilyuk-RPP2013} was attracted by recent experiments
demonstrating extremely fast ($\sim 10^{-12}$ s) magnetization
reversal triggered by a single femtosecond laser pulse in ferrimagnetic metallic rare-earth (RE) - transition
metal (TM) alloy GdFeCo.\cite{Stanciu-PRL2007,Vahaplar-PRL2009} Very recently, experimental observation of ultrafast laser-induced switching was reported
in a variety of the engineered ferrimagnetic structures, showing that this process is not specific for the RE-TM single phase alloys, but can be realized in exchange coupled RE-TM multilayers, as well as heterostructures comprised by two TM layers antiferromagnetically coupled through 0.4\,nm nonmagnetic metallic interlayers.\cite{Mangin-NatureMat2014}

Most importantly, experimental studies have demonstrated that the all-optical reversal of magnetization is not precessional and relies solely on subpicosecond quenching of the magnetizations of RE and TM sublattices.\cite{Vahaplar-PRL2009} Furthermore, as it was revealed by the time-resolved X-ray experiments and supported by the atomistic simulations,\cite{Radu-Nature2011} the laser-induced quenching of the TM and RE sublattice magnetizations occurs on distinct time-scales. As a result, the magnetization reversal proceeds via non-equilibrium transient ferromagnetic
state.\cite{Ostler-NComm2012} Such non-equilibrium dynamics allows for the deterministic magnetization reversal, without any need for other stimuli defining the magnetization direction. We note, that circularly polarized laser pulse polarization was mostly used for triggering the all-optical magnetization reversal. \cite{Stanciu-PRL2007,Vahaplar-PRL2009,Steil-PRB2011,Vahaplar-PRB2012,Mangin-NatureMat2014,Hassenteufel-AdvMater2013,Alebrand-APL2012,Alebrand-PSSA2012} However, it has recently been proposed that the difference in the magnetization reversal processes triggered by left- and right-handed laser pulses can be explained to a large extent by a magnetic circular dichroism possessed by the studied samples.\cite{Khorsand-PRL2012}

Naturally, microscopical mechanism underlying such unconventional response of magnetization of a ferrimagnetic metallic system to a femtosecond laser pulse is the subject of intense discussion nowadays. In Refs. \onlinecite{Vahaplar-PRB2012,Ostler-NComm2012,Atxitia-PRB2013,Barker-SciRep2013,Atxitia-PRB2014} atomistic and multiscale calculations based on the Landau-Lifshitz-Bloch equation\cite{Garanin-PRB1997} for the ensemble of the exchange-coupled spins have been successfully employed to account for the main features of the all-optical magnetization reversal. This approach allowed describing the all-optical reversal in both single phase alloys and exchange-coupled multilayers.\cite{Evans-APL2014} In Ref.\,\onlinecite{Kimel} comprehensive phenomenological model based on the Onsager's relations suggested by Baryakhtar \cite{Bar} was developed in order to account for the reversal via transient ferromagnetic-like state. This theoretical study introduced the exchange-dominated regime of laser-induced dynamics in a ferrimagnet, which allows the reversal of magnetization solely due to the ultrafast heating. This work highlighted the importance of the angular momentum exchange between the sublattices. Understanding microscopical processes responsible for this angular momentum exchange became, therefore, the key issue in theoretical studies of the laser-induced magnetization reversal. In Ref.\,\onlinecite{Barker-SciRep2013} the two-magnon bound state was proposed to mediate the angular momentum transfer. In Ref. \onlinecite{Wienholdt-PRB2013} dissipationless energy and angular momentum exchange between TM and RE sublattices mediated by 5$d$-4$f$ exchange coupling in RE ions was analyzed as the driving mechanism for the all-optical magnetization reversal. The role of the exchange electron-electron scattering in the magnetization reversal was recently discussed in Ref.\,\onlinecite{Baral}.

Here we consider the problem of the angular momentum exchange between two nonequivalent magnetic sublattices in a metal in the frameworks of a general microscopic model based on the rate equations. This model describes evolution of the occupation numbers of two \textit{different} ferromagnetic sublattices coupled antiferromagnetically. They are formed either by nearly localized $d$-electrons in a case of a transition metal sublattice or localized $f$-electrons in a case of a rare-earth metal sublattice.
The coupling between the sublattices is mediated by delocalized $s$- or $p$-electrons. In the frameworks of this model we demonstrate that the spin exchange between the localized ferromagnetic subsystems is mediated by delocalized electrons and is triggered by ultrafast increase of the temperature of the latter. This leads to the switching of the net magnetization without any additional stimuli, such a external magnetic field or circular polarization of light. Importantly, the model we propose is not restricted to the case of RE-TM alloys or heterostructures, but is also applicable for the case of the structures composed by two different transition metals.

Furthermore, since the laser pulse only plays a role of the stimulus supplying energy to the delocalized electrons, we consider feasibility of the magnetization reversal triggered by a short pulse of external electric bias in the ferrimagnetic system either imbedded into metallic point contact or
sandwiched between two tunnel junctions. We show that, first, this alternative approach for driving the magnetic system into the strongly-nonequilibrium
state enables one to tune the demagnetization times by
variation of the bias. This is important since the reversal
depends on a delicate interplay between demagnetization time and
cooling time of the mobile electrons. Second, in this case one deals with a compact nanoscale device compatible with existing spintronics applications.

The paper is organized as follows. In Section \ref{Sec:theory} we introduce the microscopical model describing the evolution of the ferrimagnetic metallic system in response to the rapid increase of the delocalized electrons temperature. In Section \ref{Sec:laser} we discuss the applicability of the proposed model to the process of the all-optical reversal demonstrated experimentally. In Section \ref{Sec:bias} we consider the electric bias induced reversal either in point contacts or in tunnelling structures.

\section{Theoretical model of magnetization reversal in a metallic ferrimagnet}\label{Sec:theory}

\subsection{Model of a metallic ferrimagnet}

\begin{figure}[h]
\includegraphics[width=8.5cm]{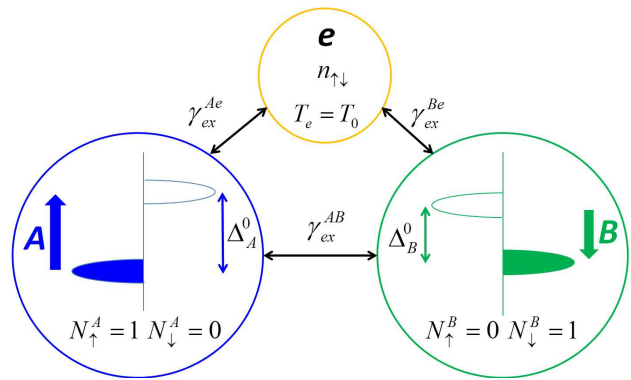}
\caption{(Color online) The A, B and $s$-subsystems comprising ferrimagnetic metal. $\Delta^0_{A,B}$ are the exchange splittings in the equilibrium. $N^{A,B}_{\uparrow\downarrow}$, $n_{\uparrow\downarrow}$ are the occupation numbers of the subsystems A, B and e. Subscripts $\uparrow,\downarrow$ denote up and down spin states with respect to the initial net magnetization direction. $\gamma_{ex}$ are the exchange constants between corresponding subsystems} \label{Fig:System}
\end{figure}

We start our consideration by introducing three interacting
electronic subsystems (Fig.\,\ref{Fig:System}). We denote two ferromagnetic
sublattices as A and B. For a sake of clarity A is the transition metal $d$-electrons subsystem, while B is either $d$-electrons or the rare-earth metal $f$-electrons subsystem. The A and B subsystems could comprise either single phase alloy or exchange-coupled layers.
The third subsystem is
the mobile $s$- or $p$-electrons (e). In the structures where both A and B sublattices are based on the TM elements
these mobile electrons do not give an important contribution to the
ferromagnetism of either of A and B sublattices. By contrast, mobile electrons support ferromagnetism of the rare-earth sublattice B via the indirect exchange with the ferromagnetic TM sublattice A. In our model these are the
mobile electrons that play a decisive role in energy and angular momentum exchange
within the sample. In particular, we assume that their coupling to
$d$- and $f$- electrons controls the energy distribution in
the corresponding subsystems. For a sake of convenience in the following discussion we consider $s$-electrons as the mobile electrons, while all the conclusions are valid for the case of mobile $p$-electrons as well.

For a sake of simplicity we consider the spin subsystems $A$ and $B$ characterized by pronounced peaks in energy distribution. Both subsystem are assumed to be strong ferromagnets and, thus, the exchange splittings $\Delta^0_{A}$ and $\Delta^0_{B}$ for these subsystems are larger than the widths of the corresponding energy peaks, as shown in Fig.\,\ref{Fig:System}. $\Delta^0_{A,B}$ describe exchange between neighboring ions comprising the subsystems A and B, and are equal to the average exchange couplings corresponding to the Weiss field. Under assumption of A and B being strong ferromagnets, the occupation numbers $N^A_\downarrow$, $N^B_\uparrow=0$, as illustrated in Fig.\ref{Fig:System}. Here the subscripts $\uparrow,\downarrow$ correspond to the up and down spin directions with respect to the net magnetization direction. Here we consider the case, when the magnetization of the A subsystem is larger than the magnetization of the B subsystem.

\subsection{Rate equations for the ferrimagnetic metallic system}

The excitation of the described above system is introduced in our model as a rapid increase of the temperature $T_e$ of the mobile electrons. We consider instantaneous increase of $T_e$ at the time $t=0$ followed
by the slow decrease, governed by the processes specific for the
system in consideration.

In order to simulate response of the A and B subsystems to the rapid increase of the temperature of the mobile electrons we exploit the fact that at temperatures close to the
critical ones the suppression of ferromagnetism of the TM sublattice A occurs mainly via the Stoner excitations which are created by a transfer of the $d$-electron from a majority band to a
minority band (Fig.\,\ref{Fig:Mechanism}(b)). Such a transfer leads to a decrease of the subsystem magnetization and is naturally related to an energy and angular momentum cost which is supplied by the mobile $s$-electrons. The decrease of magnetization of the A
subsystem is compensated by the spin reversal $\downarrow\rightarrow\uparrow$ of the $s$-electron mediating the
excitation. If the $s$-electrons are simultaneously coupled to
both A and B subsystems, they can effectively lead to spin
exchange between A and B subsystems. Thus there is the indirect interaction between total spins of A and B
subsystems, which is spin conserving in a natural way.

To describe this interaction we write down the rate
equations for the occupation numbers of sites corresponding to subsystems A ( $N^A_{\downarrow\uparrow}$), B ($N^B_{\downarrow\uparrow}$) and occupation numbers of $s$-electrons states ($n_{\uparrow\downarrow}$) participating in the exchange scattering:

\begin{widetext}
\begin{eqnarray}\label{total}
\frac{{\rm d} N^A_{\downarrow \uparrow}}{{\rm d} t} &=&
-\int\frac{{\rm d} \varepsilon}{T_e} \left[\frac{1}{\tau_{Ae}} \left(n_{\uparrow \downarrow}(1 - n'_{\downarrow\uparrow})N^A_{\downarrow \uparrow}(1 - N^A_{\uparrow \downarrow})- n_{\downarrow \uparrow}(1 - n'_{\uparrow\downarrow})N^A_{\uparrow \downarrow}(1 - N^A_{\downarrow\uparrow})\right) \right];
\label{total-A}\\
\frac{{\rm d} N^B_{\downarrow \uparrow}}{{\rm d} t} &=&
-\int\frac{{\rm d} \varepsilon}{T_e} \left[\frac{1}{\tau_{Be}} \left(n_{\uparrow \downarrow}(1 - n'_{\downarrow\uparrow})N^B_{\downarrow \uparrow}(1 - N^B_{\uparrow \downarrow})- n_{\downarrow \uparrow}(1 - n'_{\uparrow\downarrow})N^B_{\uparrow \downarrow}(1 - N^B_{\downarrow\uparrow})\right)\right];
\label{total-B}\\
\frac{{\rm d} n{\uparrow \downarrow}}{{\rm d} t} &=&
- [n_{\uparrow\downarrow}(1 - n'_{\downarrow \uparrow}) \left(\frac{1}{\tau_{eA}}N^A_{\downarrow \uparrow}(1 - N^A_{\uparrow \downarrow}) +\frac{1}{\tau_{eB}} N^B_{\downarrow \uparrow}(1 - N^B_{\uparrow\downarrow})\right)
\nonumber\\
&+& n_{\downarrow \uparrow}(1 - n'_{\uparrow\downarrow})\left(\frac{1}{\tau_{eA}} N^A_{\uparrow \downarrow}(1 -N^A_{\downarrow \uparrow}) + \frac{1}{\tau_{eB}}N^B_{\uparrow
\downarrow}(1 - N^B_{\downarrow \uparrow})\right)] +\frac{1}{\tau_s}(n_{\downarrow \uparrow} - n_{\uparrow\downarrow}).\label{total-n}
\end{eqnarray}
\end{widetext}

The r.h.s of the Eqs.\,\ref{total-A}-\ref{total-n} are the standard collision integrals of the Boltzmann equations describing the exchange scattering between three components of the electronic system. Each equation corresponds to the pair of the interacting subsystems. $n$ and $n'$ are the functions of energies $\varepsilon$ and $\varepsilon'$, respectively. The values of $\varepsilon$ and $\varepsilon'$ are connected by the energy conservation relations and include, in particular, the exchange splittings $\Delta_{A,B}$ within the subsystems A and B. Here we take into account that, in course of demagnetization the exchange splittings $\Delta_{A,B}$ differ from the equilibrium values $\Delta^0_{A,B}$.

In the Eqs.\,\ref{total-A},\ref{total-B} integration is performed only over the energy of the delocalized electrons energies. The integration over the energies of the states within the A- and B-subsystems distributions is omitted under the assumption that A and B subsystems are the strong ferromagnets. By contrast, the subsystem of mobile \textit{s}-electrons
has a broad energy distribution with Fermi energy much larger than
$T_e$. Nevertheless, Eq.\ref{total-n} is written only
for those \textit{s}-electrons which are effectively coupled to A
and B subsystems and their energy is within the band of a width
$\sim T_e$ around the Fermi level. The latter
fact means that the phase volume of the $s$-electrons involved in the exchange scattering is smaller than the phase volume
of either of the ferromagnetic subsystems A and B. Furthermore, it allows to assume the $s$-electrons densities of states within corresponding energy interval to be nearly constant. In addition, in what follows we do not not take into account the energy dependencies of $n$ as well as energy dependencies of relaxation times $\tau$.

$\tau_{A,B;e}$ are the effective electron-electron relaxation times, characterizing $A-s$ and $B-s$ exchange scattering processes. The factors $1/T_e\tau_{(A,B)e}$ in Eqs.\,\ref{total-A},\ref{total-B} are the probabilities of the exchange scattering involving $s$-electrons normalized with respect to energy $\varepsilon$. The values $1/\tau_{(A,B)e}$ are of the order of total scattering probabilities, since the integration over $\varepsilon$ is only within the energy interval $\sim T_e$.

The values $1/\tau_{e(A,B)}$ describe the probabilities of exchange scattering of $s-$electrons by $A$ and $B$ subsystems. The effective exchange scattering probability of $s$-electrons including both relaxation channels is given by $1/\tau_{ee} = 1/\tau_{eA} + 1/\tau_{eB}$.

Characteristic time $\tau_s$ describes the angular momentum exchange between the $s$-electrons and the external bath.

Eqs.\,\ref{total-A}-\ref{total-n} take into account spin balance within the system only and thus do not include processes leading to the thermal equilibrium. We assume that the characteristic times for electron-electron processes, responsible for the thermalization within considered subsystem are smaller than spin relaxation times. The evolution of the temperature $T_e$ following the instantaneous increase, is governed by electron-phonon processes and heat withdrawal, which is specific for different systems. These processes are considered to be slower than the introduced above characteristic times $\tau$ responsible for the angular momentum transfer.

Finally, we stress that in this model the effect of direct A-A and A-B or indirect B-B exchange couplings is not included. The processes involving these interactions are expected to be related to spin reversals leading to ferromagnetic or antiferromagnetic ordering in corresponding subsystems. We believe that at highly-nonequilibrium state the spin-conserving processes considered above are more efficient and fast than the ones including spin dissipation and, thus, are the dominant mechanism of the spin-redistribution. In order to set the criterion for a range of the mobile electrons temperatures $T_e$ where the exchange scattering governs the evolution of a particular ferromagnetic system subsystem, we take into account that this process is effective only for $T_e > \Delta^0_{A,B}$. Consequently, we introduce effective partial critical temperatures of the A and B subsystems, which are related to the corresponding exchange splittings $T^{A;B}_C\sim\Delta^0_{A,B}$. Strictly speaking the concept of critical temperature holds only for thermodynamical limit. In the equilibrium the critical temperatures of A and B sublattices coupled via  mobile electrons should be considered equal, in agreement with the experimental data.\cite{Ostler-PRB2011} In the strongly non-equilibrium state of the medium, if the rate of electron-electron inelastic scattering within given subsystem is higher than the rate of the corresponding sublattice magnetization evolution, one can introduce the partial electron temperature. Electron-electron inelastic scattering, responsible for the electron thermalization is, typically, in the range of 50-300\,fs.\cite{Beaurepaire-PRL1996,Hohlfeld-PRL1997,Bigot-chapter,Bovensiepen-JPCM2007} In the non-equilibrium state the A and B subsystems can be also considered as partly decoupled from each other and, therefore, we can discriminate between partial values of the critical temperatures $T^{A;B}_C$ of these subsystems. We will consider the Curie temperatures for uncoupled A (pure TM metal) and B (pure TM or RE metals) systems as the partial critical temperatures $T^{A}_C$ and $T^{B}_C$, respectively.

\subsection{Exchange scattering probabilities and relaxation times}

From the Eqs.\,\ref{total-A}-\ref{total-n} it follows that the efficiency of spin decay within a given subsystem is related to the purely spin-conserving $s-d$ or $s-f$ exchange scattering and is controlled by relaxation rates $1/\tau_{(A,B)e}$. Correspondingly, the decay rate is higher for a subsystem where this parameter is larger, i.e. the exchange coupling with mobile $s$-electrons is stronger. One expects that the exchange scattering between $s$-electrons and the corresponding ferromagnetic subsystem is more pronounced if the latter possesses strong exchange interaction within itself. Although the evolution of magnetization in any of the subsystems includes not only spin transfer between A or B subsystem and s-subsystem, controlled by $1/\tau_{(A,B)e}$ but also the spin decay within $s$-subsystem ($~1/\tau_s$), we expect that it is the difference between the values of $\tau_{(A,B)e}$ that leads to the distinct times of spin decay within A and B subsystems.

Numerical estimation of the relaxation rates $1/\tau_{(A,B)e}$ and $1/\tau_{e(A,B)}$ requires knowledge of the electron spectrum of all involved systems. Here we use simplified estimates. As it is known, for electron-electron scattering in standard metals relation $1/\tau\sim T_e^2/\varepsilon_F \hbar$ holds, where $\varepsilon_F$ is the Fermi energy (see, e.g., Ref.\,\onlinecite{Abrikosov}). This expression makes use of the momentum conservation law for the electron system where $T_e<<\varepsilon_F$. In the case considered here the situation is different since the electron scattering takes place between two different electronic subsystems, with one of them (A or B) characterized by very narrow energy band, and, thus, the effective mass of electrons in which is much larger than in the subsystem of the mobile electrons. In this case the momentum conservation law can be disregarded and thus the electron-electron scattering time has a form close to the one for electron scattering by impurities. Therefore, one can estimate the relaxation time as  $1/\tau\sim\sigma n_\mathrm{eff} v_{r}$ where $\sigma$ is the scattering cross-section, $n_\mathrm{eff}$ is the effective concentration of scatterers, and $v_{r}$ is the relative velocity of scattered electron with respect to the scatterer (see, e.g., Ref.\,\onlinecite{Abrikosov}). Taking these considerations into account, one obtains for the characteristic time of the exchange scattering of the A or B subsystem electrons by the $s-$ electrons with the spherical Fermi surface:
\begin{equation}
\frac{1}{\tau_{(A,B)s}}\sim\gamma^{(A,B)e}_{ex}\frac{\hbar ^2 n T_e}{\varepsilon_F^{3/2} m^{3/2}},\label{estimeates-As}
\end{equation}
where $n$ is the concentration of the $s$-electrons, $\varepsilon_F$ is the Fermi energy of the $s$-electrons, $m$ is their mass. $\gamma^{(A,B)e}_{ex}$ is the dimensionless $A-s$ and $B-s$ exchange constant, which absorbs the dependence of the exchange scattering probability on the exchange splittings $\Delta^0_{A,B}$. Here the estimates $\sigma_{(A,B)e}\sim\hbar^2/m\varepsilon_F$,  $v_{r}\sim\varepsilon_F^{1/2}/m^{1/2}$ , $n_{eff}\sim n T_e/\varepsilon_F$ are used.

In its turn, for the probability of exchange scattering of $s-$electrons by the ones of A or B subsystems we obtain for the case $T_e > \Delta_{A,B}$:
\begin{equation}
\frac{1}{\tau_{s(A,B)}}\sim\gamma^{(A,B)e}_{ex}\frac{\hbar^2 N^{A,B}  }{\varepsilon_F^{1/2} m^{3/2}},\label{estimeates-sA}
\end{equation}
where $N^{A,B}$ is the concentration of the subsystem A or B.
Thus, according to these expressions, the values of $\tau_{(A,B)s}$ and $\tau_{s(A,B)}$ are different for the same value of the exchange scattering crossections and exchange constants between the A(B) subsystem and the mobile electrons.

As Eqs.\,\ref{estimeates-As},\ref{estimeates-sA} demonstrate, the relaxation probabilities $1/\tau_{(A,B)e}$ and $1/\tau_{ee} = 1/\tau_{eA} + 1/\tau_{eB}$ possess different temperature dependencies. For the exchange scattering time $\tau_{(A,B)e}$ we have $1/\tau_{(A,B),e}\propto T_e$. The exchange scattering time $\tau_{ee}$ is, in turn, temperature -independent. This is in contrast to standard electron-electron scattering where $1/\tau\sim\hbar T_e^2/\varepsilon_F$. The later relation holds for electron-electron scattering between $s$-electrons responsible for thermalization at the initial stage.

\subsection{Evolution of the ferrimagnetic system with nearly quenched sublattice magnetizations}

Using Eqs.\,\ref{total-A}-\ref{total-n} we consider the evolution of the magnetizations of the A and B sublattices triggered by an instantaneous increase of the temperature of the mobile electrons in the metallic ferrimagnet. If one would deal with a single ferromagnetic subsystem, e.g. A, coupled to the mobile electrons, the rapid increase $T_e>T^C_A$ would trigger the decrease of the sublattice magnetization, i.e. decrease of $N^A_\uparrow$ and increase $N^A_\downarrow$ due to spin transfer to the mobile electrons via exchange scattering and the following spin decay within $s$-subsystems. This would finally lead to total suppression of magnetization.

It is important to stress that the values of $\Delta_{A,B}$ decrease in the course of the demagnetization process. In particular, when the average magnetization of the A sublattice tends to zero, the same holds to the average exchange fields. As a result, in the mean field approximation $\Delta_A$ tends to zero as well. However, locally, given magnetic ions from the A sublattice is exchange coupled to the nearest neighboring ions. The distribution of local magnetic moments does not possess any long range order, and the sum magnetization of the neighboring ions fluctuates depending on the spatial position. Since in the equilibrium $\Delta_A^0\propto\mathcal{N}_A$, where $\mathcal{N}_A$ is the number of neighbours, the mean exchange splitting $\Delta_A\sim\Delta^0_A/\mathcal(N)^{1/2}_A$ when the average magnetization of the A sublattice is zero.

For the antiferromagnetically coupled A and B subsystems the evolution of their magnetizations is somewhat more delicate than in the case of the single sublattice system. The exchange scattering with the mobile electrons leads to the redistribution of the total spin between the subsystems according to the factor $\tau_{A,e}/\tau_{B,e}$. The suppression of total magnetization would occur only via the spin non-conserving
process, which is described by the term $(n_{\downarrow\uparrow}-n_{\uparrow\downarrow})/\tau_s$ in Eq.\,\ref{total-n}. Without this term the total suppression of magnetization in the system is possible only if the magnetizations of A and B subsystems are equal initially.

\begin{figure}[h]
\includegraphics[width=8.5cm]{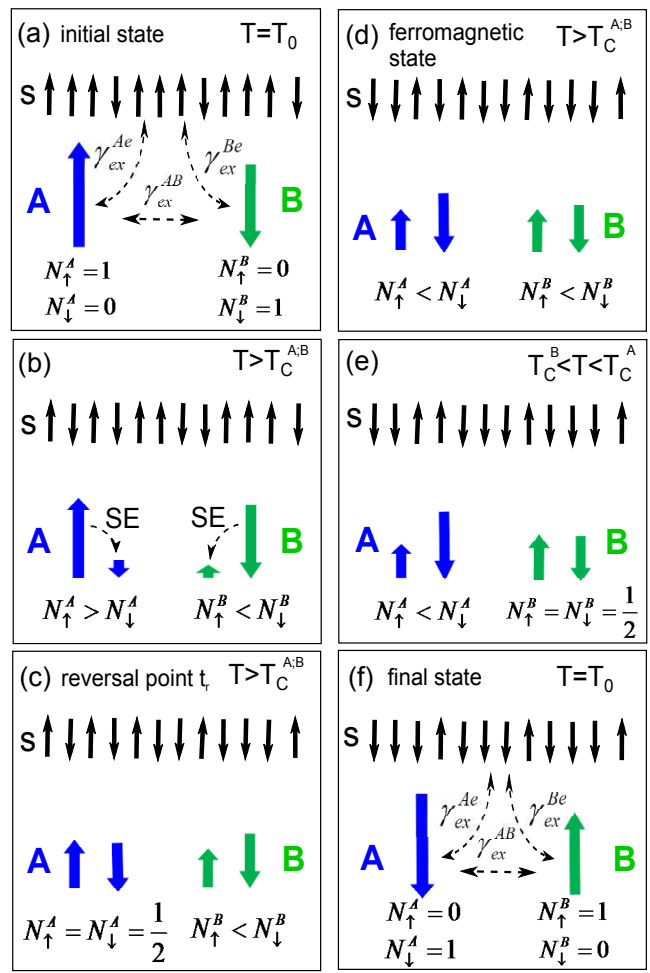}
\caption{(Color online) Evolution of the A, B and $s$-subsystems
following the rapid increase of $T_e$. Snapshots of the subsystems at six distinct time moments are shown: (a) the initial state; (b) Stoner excitations (SE) decrease the magnetization of the A and B subsystems; (c) reversal point $t_r$, where the magnetization of the A subsystem vanishes; (d) the magnetization of the subsystem A is partly restored in a new direction, while the magnetization of the B subsystem is remains finite - transient ferromagnetic-like state; (e) the magnetization of the A subsystem is restored sufficiently for reversing the magnetization of the B subsystem via the direct A-B exchange interaction; (f) the final state. Corresponding temperatures of the mobile electrons subsystem with respect to the partial temperatures $T^{A;B}_C$ are shown for each snapshot. See text for details.} \label{Fig:Mechanism}
\end{figure}

In order to illustrate the evolution of the magnetizations of the A and B subsystems which follows the instantaneous increase of the $s-$electrons temperature $T_e>T_C^{A,B}$ we consider the situation when the creation of Stoner excitations
completely suppresses magnetization of the one of the subsystems. We consider the case, when $\tau_{A,e} < \tau_{B,e}$, which is consistent with the notations we accommodated, i.e. A and B subsystems are formed by $d-$ and $f-$electrons, respectively. Then magnetization of the A sublattice is quenxhed, $N^A_\uparrow=N^A_\downarrow$, while the magnetization of the B subsystem remains finite, $N^B_\downarrow>N^B_\uparrow$ (Fig.\,\ref{Fig:Mechanism}(c)). In this case the rate equation for the A subsystem takes a form:
\begin{equation}\label{dNAdt}
\frac{{\rm d} N^A_{\downarrow \uparrow}}{{\rm d} t} |_0
=\frac{1}{4\tau_{Ae}}(n_{\downarrow \uparrow} - n_{\uparrow
\downarrow}).
\end{equation}
Here the notation $|_0$ means the configuration where $N^A_{\downarrow} = N^A_{\uparrow}$. The further calculations give that at $N^A_{\uparrow}
 \simeq N^A_{\downarrow} \simeq 1/2 $, $|n_{\uparrow} - n_{\downarrow}| << n_{\uparrow}$ the rate equation of the $s$-electrons takes a form:
\begin{widetext}
\begin{equation}\label{dndt}
\frac{{\rm d} n_{\uparrow \downarrow}}{{\rm d}t} =
\frac{1}{\tau_s}(n_{\downarrow \uparrow} - n_{\uparrow
\downarrow}) + \frac{1}{4\tau_{ee}}(n_{\downarrow \uparrow} -
n_{\uparrow \downarrow}) + \frac{1}{4\tau_{Be}}(N^B_{\uparrow
\downarrow} - N^B_{\downarrow \uparrow}) + \frac{1}{4\tau_{Ae}}
(N^A_{\uparrow \downarrow} - N^A_{\downarrow \uparrow}),
\end{equation}
\end{widetext}
Here we assumed $n_{\uparrow} \sim n_{\downarrow} \sim 1/2$.

Then we recall that the phase volume of the $s$-subsystem is smaller than the phase volumes related to the subsystems A and B. Therefore, the characteristic time of evolution of the magnetic system as a whole is much larger than the characteristic relaxation times $\tau_{eA}$, $\tau_{eB}$, $\tau_s$, relevant for the mobile electrons. Consequently, we neglect time derivative in the l.h.s. of Eq.\ref{dndt}. This leads to the expression
\begin{equation}\label{n}
(n_{\downarrow \uparrow} - n_{\uparrow \downarrow}) \simeq
-\frac{(\tau_{Be})^{-1}}{(4/\tau_s) + (1/\tau_{ee})}(N^B_{\uparrow
\downarrow} - N^B_{\downarrow \uparrow}).
\end{equation}
Since at the time moment when $N^A_{\downarrow} =
N^A_{\uparrow}$ the magnetization of the subsystem B is non-zero and the r.h.s. of Eq.\ref{n} is positive, Eqs.\,\ref{n},\ref{dNAdt} show that the magnetization of the subsystem A
changes its sign. Therefore, after this moment the total configuration of
the A-B system starts to be {\it ferromagnetic} ((Fig.\,\ref{Fig:Mechanism}(d))). Note that this
happens in course of decay of total magnetization of the system, provided that spin non-conserving processes described by $\tau_s$ are taken into
account. We denote the moment of time, corresponding to reversal of the magnetization of the A subsystem, as the "reversal point" $t_r$. It is important to note that at the times $t>t_r$ the former minority electrons of the A subsystem become to be majority and \textit{vice versa}.
Correspondingly, the reference for the Stoner excitations is
changed - now they are referred with respect to the "new" orientation
of magnetization. Therefore, the exchange with B subsystem via
$s$-electrons leads to a {\it decrease} of the excitations number in the A subsystem.

If the electron temperature $T_e$ would be kept constant after $t_r$, the magnetizations in both subsystems would vanish, provided finite $\tau_s$. Since $T_e$ gradually decreases after $t=t_r$ at some moment it reaches the critical
temperature $T^A_C$, while is still above $T^B_C$ (Fig.\,\ref{Fig:Mechanism}(e)). At this moment, as shown above, the magnetization of the A subsystem is non-zero
and is aligned along the initial magnetization direction of the subsystem B. Then, the presence of a gap between new majority and minority
bands in the A subsystem is restored, and the electron-electron $s-d$ exchange scattering can not support anymore some pairs of the Stoner excitations with the "new" reference. This leads to the increase of the magnetization of the A subsystem aligned to the
direction of initial magnetization of B subsystem. We note that,
simultaneously, the inter-A exchange interactions are also
restored. To the contrary, magnetization of the B subsystem, characterized by the smaller critical
temperature $T^B_{C}$ continues to decrease due to the Stoner excitations supported by the $s$-electrons subsystem, according to Eqs.\ref{total-A}-\ref{total-n}.

When the electron temperature $T_e$ decreases down to the value $T^B_C$ the subsystem A already acquired the magnetization large
enough to force the subsystem B to reconstruct its magnetization state according to the new magnetization state of the subsystem A via indirect antiferromagnetic exchange (Fig.\ref{Fig:Mechanism}(f)).

The critical value of $T_e$ corresponding to irreversible switching can be estimated from $\Delta^0_A$ or $T^C_A$. At this critical value of $T_e=T^C_A$ the self-consistent character of exchange is restored and the standard Weiss field is formed. Such an estimate is mostly a semi-qualitative one since the process of transition from strongly non-equilibrium regime to an equilibrium has a complex character.It could be calculated with a help of numerical methods provided one has a detailed information concerning temperature behavior of $\tau_s$, the heat withdrawal processes etc.

We would like to emphasize that at the strongly non-equilibrium state considered above the main processes defining the spin kinetics within the system are related to \textit{spin-conserving} exchange scattering. With lowering the temperature below critical temperatures $T^{A,B}_C$ of the subsystems A and B this scattering becomes suppressed, while the standard exchange interactions implying \textit{spin non-conserving} processes start to play the decisive role. However, close to the reversal point $t_r$ the direct A-B exchange is still suppressed since it is proportional to the product of total spins of the A and B subsystems. This explain the fact that, while the demagnetization of subsystems A and B takes relatively short time below 1\,ps, the total equilibrium is established at times of tens of picoseconds.\cite{Vahaplar-PRL2009}

Finally, we would like to note that, despite the assumptions and simplifications made above, we expect that the main conclusions drawn on the base of our simplified model hold also for more realistic ferromagnetic systems with smaller exchange splitting and for more complex nature of the demagnetization process. Indeed, in our model the Stoner excitations are considered to be the mechanism of the quenching of magnetization of the sublattices A and B. The physical picture of the demagnetization process is more complicated and may include spin fluctuations,\cite{Carpene-PRB2015} arising e.g. from magnons. Magnons were recently suggested to mediate the angular momentum transfer in the course of all-optical magnetization reversal.\cite{Barker-SciRep2013} Qualitatively, the scheme we applied could be used to include these types of excitations in addition to the Stoner's ones as the mechanism of quenching of magnetization of the sublattices. These excitations could be magnons of different types including both $A$ and $B$ subsystems. However, under assumption of narrow distribution of energies within $A$ and $B$ subsystems, the relevant magnons should have specific frequencies satisfying the energy conservation. Since the subsystems $A$ and $B$ are different, the magnons coupled to these subsystems have different frequencies, and thus can not simultaneously be coupled to $A$ and $B$ subsystems and can not transfer spin between the two subsystems while this transfer is an important ingredient of the reversal process. Therefore, in order to include in the suggested model the magnons mediating the angular momentum exchange, one cannot use anymore the simplification of narrow distribution of energies within $A$ and $B$ subsystems.

Another simplifications included to our model is that both subsystems A and B are considered to have spin 1/2, since only two projections of spins are considered in Eqs.\ref{total-A}-\ref{total-n}. Thus in this case we neglect a possible role of intermediate spin projections in the process of demagnetization of the A and B sublattices. Strictly speaking, it does not hold for rare-earth ions. Nevertheless we believe that our picture gives at least a qualitative explanation of the effect in a general case. Indeed, in any case the switching between the two magnetization orientations imply transition between the two extreme spin projections and, correspondingly, the gap between the two projections can be considered as $\Delta^0_{A,B}$.

\section{Magnetization reversal induced by a femtosecond laser pulse}\label{Sec:laser}

The ultrafast all-optical switching based on the
effect of exchange electron-electron scattering was recently
discussed in Refs.\,\onlinecite{Baral,Gridnev-PRB2013}.
According to Fig.\,2 of Ref.\onlinecite{Baral}, the possibility of
magnetization switching in the model is mainly related to the
following. (i) The total numbers of localized and itinerant spins were
equal. (ii) The spin of localized electrons was twice as large as
the spin of the itinerant carrier. In this case the exchange
scattering at a high level of excitation can force the average spin of the
itinerant electrons to the opposite direction dictated by the
local spins to the extend sufficient for the $d-d$ exchange to "fix" this "new" direction in the course of the electron
cooling. The weaker exchange between localized and itinerant carriers in this case can
only affect the direction of local spins. However, to our opinion, this scenario can not be applied to the case of
multilayered systems,\cite{Mangin-NatureMat2014} since the penetration length of the $d$-electrons through such structures including interlayers is questionable.

In Ref. \onlinecite{Gridnev-PRB2013} the \textit{sp-d} model of ultrafast demagnetization\cite{Cywinski-PRB2007} was applied in order to describe the all-optical magnetization reversal scenario in a ferromagnet. In this case, however, the optical pumping resulting in itinerant spin polarization had to be included in the model in order to facilitate the magnetization reversal.

We also note that dissipationless energy and angular momentum exchange between TM and RE sublattices as the driving mechanism for the optical magnetization reversal has been explored in Ref.\,\onlinecite{Wienholdt-PRB2013}. There the 5$d$-4$f$ exchange coupling in RE ions was proposed to be decisive for the reversal, which naturally limits the applicability of the model to the case of magnetization reversal in RE-TM alloys and heterostructures, but not to the engineered ferrimagnets based on transition metals solely.

Now we compare the magnetization reversal scenario given by our model with the experimental findings reported in the papers on ultrafast all-optical magnetization reversal. When the metallic film is subjected to a intense femtosecond laser pulse the temperature of the mobile electrons in the film is increased on a time-scale of the laser pulse.\cite{Beaurepaire-PRL1996} Therefore, when describing the interaction of laser pulses with
metal, one typically introduces a rapid increase of the
temperature $T_e$ of the "bath" formed by the mobile electrons.\cite{Beaurepaire-PRL1996,Ostler-NComm2012,Evans-APL2014}
Estimates based on the heat capacity of mobile electrons in a
transition metal give that the 100\,fs laser pulse of
$\sim$1\,mJ/cm$^2$ fluence leads to the increase of $T_{e}$ up to
$\sim$1200\,K.\cite{Vahaplar-PRB2012}. This is followed by the
decrease of $T_{e}$ governed by the electron-lattice relaxation
times. Calculations in frames of the two-temperature
model\cite{Kaganov-JETP1957} performed e.g. in
Ref.\,\onlinecite{Vahaplar-PRB2012} yielded that the temperature of
the electronic bath in a conducting metal equilibrates with the
lattice one on a time scale of a few picoseconds. On this time
scale $T_{e}$ drops to the values comparable to the equilibrium
Curie temperature of a medium, with exact value being determined
by the amount of energy transferred to the system by the laser
pulse.\cite{Kazantseva-EPL2008} Therefore, the evolution of the
electronic temperature during and after the excitation by the
femtosecond laser pulse corresponds well to the conditions implied
in our model.

The magnetization reversal scenario yielded by the proposed
model agrees well in the main details to the results of
various experiments aimed at revealing the the underlying
mechanism of the all-optical magnetization reversal in
ferrimagnetic RE-TM alloys.

In particular, studies of the laser-induced dynamics of
magnetizations of Fe and Gd sublattices in GdFeCo alloy have shown
that they demagnetize on different time scales on 140 and 400\,fs
respectively.\cite{Radu-Nature2011} This results in occurrence of
the transient ferromagnetic-like state, which is described
by Eqs.\ref{dNAdt}-\ref{n} in the model considered in Section\,\ref{Sec:theory}. We note, that the numerical
calculations reported in Ref.\,\onlinecite{Radu-Nature2011}
suggested that there is a proportionality between the
demagnetization times observed for the TM and RE sublattices and
corresponding ratios between their magnetic moments and Gilbert damping
constants $\mu_\textrm{TM;RE}/\lambda$. In our model distinct
demagnetization times are accounted by distinct exchange relaxation rates $1/\tau_{Ae}>1/\tau_{Be}$, which are correlated to the exchange splittings $\Delta^0_{A,B}$ possessed by the A and B subsystems.

In the presented model the magnetization reversal is driven by the
exchange of angular momentum between the A and A sublattices
mediated by the mobile electrons, while the transfer of the
angular momentum to other reservoirs (lattice) is only responsible
for overall decay of magnetization of the whole system. Earlier,
it has been suggested, based on the studies of the ultrafast
laser-induced demagnetization in GdFeCo alloys, that the angular
momentum transfer from TM to RE sublattice plays an important role
in the process.\cite{Medapalli-PRB2012} Spatially- and element-resolved studies of the reversal dynamics in GeFeCo have shown that there is the angular momentum transfer between Gd-rich and Fe-rich nanoscale areas in the GdFeCo sample which accompanies the reversal.\cite{Graves-NMater2013} Recently, has been reported that there is a transfer of the angular momentum between RE
and TM sublattices of the metallic
ferrimagnetic alloys CoGd and CoTb induced by the action of the laser pulse and monitored by the spin- and orbital-resolved X-ray technique.\cite{Bergeard-NatureComm2014} The results of the element-specific studies of the laser-induced demagnetization and reversal in TbCo alloys\cite{Alebrand-PRB2014} supported further the involvement of the exchange between the RE and TM sublattices in these processes.

When introducing our model we did not specify whether the A-e-B ferrimagnetic system should be single phase one or comprised by coupled A and B layers.
Thus, we argue that this model accounts well for the results reported in Ref.\,\onlinecite{Mangin-NatureMat2014}, where the all-optical reversal was observed in four distinct types of single-phase and multilayered synthetic metallic ferrimagnets.

Therefore, the model considered here captures the general picture of the laser-induced magnetization reversal in a metallic ferrimagnet. However, due to a number of simplifications applied and since our model does not include the realistic band structure of a ferrimagnetic metal, it cannot account for a number of experimental evidences, which we discuss below.

Ongoing studies of the magnetization reversal reveal very diverse and even contradictory features of the process in the RE-TM metallic alloys of various compositions. The important issue of the role of the magnetization compensation point possessed by ferrimagnets has been studied experimentally in both alloys\cite{Stanciu-PRL2007,Vahaplar-PRB2012,Ostler-NComm2012,Medapalli-EPJ2013,Alebrand-APL2012} and engineered multilayer structures.\cite{Mangin-NatureMat2014} Number of experiments have demonstrated,\cite{Vahaplar-PRB2012,Ostler-NComm2012,Medapalli-EPJ2013} that the reversal can be realized for ferrimagnets which equilibrium temperature either below compensation point or above it, which agrees well with the proposed model. On the other hand, experiments reported in Ref.\,\onlinecite{Alebrand-APL2012} suggest that for the reversal it is essential that the compensation point is above the equilibrium sample temperature. The recent study of the reversal in the series of specially engineered ferrimagnetic structures showed that this condition holds for the majority but not for all structures.\cite{Mangin-NatureMat2014} Despite of these controversies, all the studies confirm that the reversal does not occur in TM-RE alloys, which are either TM-rich or RE-rich. Casting the light on this problem, in Ref.\,\onlinecite{Hassdenteufel-PRB2015} the importance of the low remanence, possessed by ferrimagnets in a vicinity of the compensation point, has been revealed. This is in agreement with the earlier reported data,\cite{Vahaplar-PRB2012} showing that the closer the sample to its compensation point, the less laser fluence is required for the reversal. Our model does not treat such details of the ferrimagnetic metal as the equilibrium ratio between the sublattice magnetizations and therefore it cannot account for the role of the magnetization and angular momentum compensation points.

Another issue which is still to be comprehended is the laser pulse duration required for the reversal. In Ref.\,\onlinecite{Vahaplar-PRB2012} the reversal in GdFeCo alloys of certain compositions could not be realized by the pulses longer than 1.7\,ps, while in Ref.\,\onlinecite{Alebrand-APL2012} the reversal by the laser pulses as long as 10\,ps was reported. Based on the present knowledge, the  reversal scenario treated in the frameworks of our model requires femtosecond laser pulses which could bring the RE-TM alloys, studied in the reported experiments,\cite{Vahaplar-PRB2012,Alebrand-APL2012} to the highly non-equilibrium state on the time scale comparable to the relaxation times $\tau_{AE}<\tau_{BE}<$1\,ps. We note, however, that the maximal pulse length required for the reversal is dependent on the balance between the rate and the degree of the electronic temperature increase, the exchange relaxation times $\tau_{Ae}$, $\tau_{Be}$ and the rate of the energy and angular momentum withdrawal $\tau_s$. Therefore, the knowledge of the details of the spin-conserving and spin-nonconserving relaxation processes in a particular ferrimagnetic samples for the particular pulse durations is essential for understanding the restrictions on the maximal pulse duration. We are not aware about the time-resolved studies of the laser-induced reversal by pulses longer than 100\,fs.

Finally, we note that recently the switching effects for purely {\it ferromagnetic} structures were reported.\cite{Lambert} In this case only the laser-pulse helicity dependent switching has been reported, reopening the discussion about the role of the light polarization open. We believe that further experimental studies, clarifying this issues are required before any conclusions regarding the mechanism of the reversal in the ferromagnets can be evaluated.

\section{Magnetization reversal induced by an electric bias pulse}\label{Sec:bias}

\begin{figure}[h]
\includegraphics[width=8.5cm]{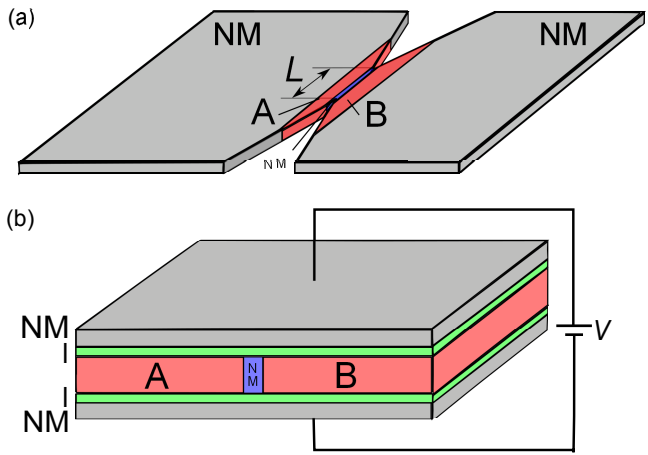}
\caption{(Color online) (a) Structure formed by the two ferromagnetic layers, A and B, separated by normal metal interlayer NM, which is
imbedded into point contact on the base of normal metal NM. The interlayer NM is chosen in a way that it supports antiferromagnetic coupling between the layers A and B. (b) Structure formed by two ferromagnetic islands A and B, separated by normal metal interlayer NM,
imbedded between two normal metal electrodes NM and separated from them by two tunneling interlayers I. The normal metal interlayer NM
supports antiferromagnetic coupling between A and B.} \label{fig2}
\end{figure}

According to the model described in Sec.\,\ref{Sec:theory} the rapid heating of the mobile electrons is sufficient for triggering the magnetization reversal. Therefore, we suggest that an electric bias pulse used as an alternative to femtosecond optical pulses and can drive the ferrimagnetic metallic system into strongly-nonequilibrium state, where the magnetization reversal can be realized. We consider a possibility of switching within the A-e-B system formed either the two ferromagnetic layers $A$ and $B$ or by the A-B metallic alloy imbedded within the metallic point contact (see Fig.\,\ref{fig2}(a)).

Again, for a simplicity we consider the model of strong ferromagnets
where the minority spins do not exist in the equilibrium (Fig.\,\ref{Fig:System}). Furthermore, we
neglect the energy distribution of both spin subsystems
thus reducing A and B subsystems to the two spin sublevels separated by the exchange energy $\Delta^0_{A}$ and $\Delta^0_{B}$, respectively. The energy distribution of the $s$-electrons is controlled by a voltage applied
to the point contact. Namely, if the width of the contact $L$ is
smaller than the diffusive length with respect to energy
relaxation, the distribution of the $s$-electrons is formed as a mixture
of electrons coming from the left bank of the contact and those coming from
the right bank, and is controlled by corresponding chemical potentials.
In the center of the contact the distribution $F$ has a
double-step form:\cite{Shkor}
\begin{equation}\label{pc}
F =\frac{1}{2}\left(F_0(\varepsilon_F-eV/2)+F_0(\varepsilon_F+eV/2)\right),
\end{equation}
where $V$ is the applied voltage, $\varepsilon_F$ if the Fermi energy, and $F_0(\varepsilon)=[1+\exp(\varepsilon-\varepsilon_F)]^{-1}$. At some distance from the center to the left or to the right the weight of the corresponding "left" or "right" contribution increases and at large distances the equilibrium distributions of
"left" and "right" types are restored. Note that the
distribution (\ref{pc}) holds near the center of the contact even for
diffusive transport provided that inelastic mean free path is smaller
than the size of the contact $L$.

As it is seen, the energy $eV$, defined by the applied voltage $V$, can play a role of effective temperature of the $s$-electron
system, and, therefore, trigger the magnetization dynamics described in Sec.\,\ref{Sec:theory}. In contrast to the case of all-optical reversal, in this case following the excitation the non-equilibrium spin occupations of $s$-electrons decay due to ballistic or diffusive transport from the contact to the banks. Thus in this case the spin relaxation time $\tau_s$ in Eq.\,\ref{n} is the escape time, which is defined as
\begin{equation}\label{escape}
\tau_s=\tau_{esc}^\mathrm{b}\sim\frac{L}{v}; \tau_s=\tau_{esc}^\mathrm{d}\sim\frac{L^2}{D},
\end{equation}
for the case of ballistic and diffusive transport, respectively. Here $L$ is the characteristic size of the contact, $v\sim10^8$\,cm/s is the electron velocity, and $D$ is the diffusion constant.

In this point-contact device one can control
both the excitation intensity (by the value of the bias $V$) and
the parameters of the excitation pulse (including the switching-on/off times). The switching-on time - if small
enough - is of no great importance. In contrast, the switching-off time should
be comparable with the time scale of the magnetization reversal. The latter, as we discussed in Sec.\,\ref{Sec:theory}
is controlled by electron-electron
exchange relaxation times $\tau_{(A,B)e}$, which are expected to be of the order of
$10^{-12}$s. Importantly, as we discuss below, the time required for the magnetization
reversal $t_r$ can be increased both by the choice of the bias
and by a proper position of the layers with respect to
the contact center, since an increase of the corresponding distance
decreases the phase volume of the electron-electron scattering
and, thus suppressing the magnetization reversal.

Let us consider an effect of the $s$-electron distribution given by
Eq.\ref{pc} on the ferromagnetic layer A imbedded into the contact
near its center. We consider the case of zero equilibrium temperature, $T_0=0$. If $eV > \Delta^0_{A}$, than the occupation $N^A_\downarrow$ of the minority level of the subsystem A is described by an equation\cite{me}
\begin{equation}\label{PC}
\frac{{\rm d} N^A_\downarrow}{{\rm d} t} + \frac{\tau_{Ae}^{-1}}{2\Delta_A}(\Delta_A
+ eV) N^A_\downarrow = \frac{\tau_{Ae}^{-1}}{4\Delta_A}(eV - \Delta_A).
\end{equation}
Now we take into account that the
occupation of the minority level leads to a decrease of the exchange field, i.e.
$\Delta_A = (\alpha/2) (N^A_\uparrow -N^A_\downarrow)$, where $\alpha$ is the proportionality factor. As we discussed above, although the average exchange splitting vanished at  $N^A_\uparrow=N^A_\downarrow$, the exchange splitting of the given ion is $\Delta_A\propto\mathcal{N}_A$ due to local fluctuations. Then $N^A_\uparrow -N^A_\downarrow = 1 - 2N^A_\downarrow$ and thus $\Delta_A = \Delta^0_A - \alpha N^A_\downarrow$, where $\alpha = 2\Delta^0_A$.

If we denote $eV = 2\Delta^0_A + \delta$, then Eq.\ref{PC} can be rewritten in a form:
\begin{eqnarray}
  \frac{{\rm d} N^A_\downarrow}{{\rm d} t} &=& - \frac{\tau_{Ae}^{-1}}{4\Delta^0_A(1 -2N^A_\downarrow)} \nonumber\\
  &\times& \left( 4\Delta^0_A (N^A_\downarrow - 1/2)^2 + 2\delta(1/2 - N^A_\downarrow)\right). \label{PC1}
\end{eqnarray}
For $\delta < - \Delta^0_A$, i.e $eV <
\Delta^0_A$, there is no non-zero solution of the  Eq.\,\ref{PC1}. The occupation of the minority level
starts naturally at the threshold value $eV = \Delta^0_A$, or $\delta = - \Delta^0_A$. The
gradual increase of $N^A_\downarrow$ with an increase of $eV$ terminates at the
value $N^A_\downarrow = 1/2$ which is reached at $\delta = 0$. Thus one
concludes that the transition to the ferromagnetic-like state with parallel A and B magnetizations takes
place at $eV = 2\Delta^0_A$. For $eV > 2\Delta^0_A$ the steady
state of the layer A corresponds to normal metal.

Close to the critical value of the bias $V_c =
2\Delta^0_A/e$, the evolution of $N^A_{\uparrow\downarrow}$ ($N^B_{\uparrow\downarrow}$) and of the exchange
field $\Delta_A$ ($\Delta_B$) with time is controlled by the difference $V - V_c$, since
the non-vanishing part of r.h.s. of Eq.\ref{PC1} scales with this difference. In the vicinity of $N^A_\downarrow = 1/2$ the difference $1/2 - N^A_\downarrow$
can be considered as a variable. In
the r.h.s. of Eq.\ref{PC1} the leading term is linear in $1/2 -N^A_\downarrow$,
and the coefficient at this term gives the rate of the
evolution. Such a slow evolution
is expected, roughly speaking, in the case when $(1/2 - N^A_\downarrow) < |eV - eV_c|/\Delta^0_A$. Correspondingly, the evolution of ferromagnetic order parameter near critical
point is defined by a characteristic time
\begin{equation}\label{tau}
t_r\sim \frac{\tau_{Ae}\Delta^0_A}{|eV - eV_c|}.
\end{equation}
Thus the time required for suppression the ferromagnetic state of the sublattice A can be tuned
by a proper choice of the bias.

Now we would like to note that a specific feature of the point
contact is a possibility to apply voltage pulses with a sharp
form. As a result, one can operate the device in a threshold way
which was demonstrated above. Thus we believe that our predictions
including Eq.\ref{tau} hold also for more realistic models
including finite energy width of the ferromagnetic
subsystem.

Another design of the bias-controlled switching device can be based on the
tunnel junctions (Fig.\,\ref{fig2}(b)). Namely, we assume that
the ferrimagnetic structure $A-s-B$ discussed above is fabricated on the
base of thin film technology, and the corresponding thin film
device is sandwiched between two tunnel junctions. Note that for
the case of the device formed by two ferromagnetic region
separated by normal metal all three components are considered to
be fabricated within the same plane, as shown in Fig.\,\ref{fig2}(b).

In this case the external bias is applied to external metallic electrodes of the tunnel junctions. If the transparency coefficient, assumed to be the same for both tunneling barriers, is $k$, than the effective time spent by a non-equilibrium electron within the device, which is the measure of $\tau_s$, is
\begin{equation}\label{escape-barrier}
\tau_s=\tau_d\sim\frac{d}{v_F}k^{-1},
\end{equation}
where $d$ is a thickness of the film forming the
device while $v_F$ is the Fermi velocity within
the metal structure. Thus we conclude that if $\tau_d$ is comparable to the characteristic electron-electron relaxation time and is larger than the characteristic electron-phonon relaxation time, then the
electron distribution function within the device is given in a
same way as in Eq.\ref{pc}.

We note that for such a design the picture is to some extent similar to the one corresponding to
optical excitation In particular, here we also deal with vertical
geometry of excitation. Then, in contrast to the point-contact design (Fig.\,\ref{fig2}(a)),
here we have no limitation in horizontal size of the
device. The important difference is that in the tunnel junction-based device
the sharp form of the electron distribution allows effective
control of the switching process by controlling the form of
the bias pulses.

The point-contact or tunnel junctions scheme where the reversal is driven by the electric bias allows to avoid a number of drawbacks which are often considered as the limitations for the all-optical reversal. As it follows from the
considerations in the Sec.\,\ref{Sec:theory}, the magnetization reversal depends on a balance between magnetization decay time $\tau_{A,B;e}$ in
$T_e \rightarrow \infty$ limit and the decay time of the electron temperature $T_e$. In a case
of optical excitation achieving such a balance may require a delicate
choice of the pulse duration and pulse
intensity and sample characteristics.\cite{Vahaplar-PRB2012} Furthermore, bringing the
optical excitation to the nanoscale is a challenging
task.\cite{Savoini-PRL2013,LeGuyader-NatureComm2015} An additional
factor which requires accurate control for application of the all-optical
magnetization reversal is related to a rate of the cooling
time, controlled by a heat withdrawal from the laser-excited spot. Thus, in
Ref.\onlinecite{Vahaplar-PRL2009}, where the record-short all-optical
write-read time of $\tau_\mathrm{w-r}$=30\,ps has been reported,
in experiments the residual heating resulted in only 83\,\%
magnetic contrast restored at $\tau_\mathrm{r-w}$. In the case of the bias driven magnetization reversal, better control of the decay of the electron temperature and the cooling can be achieved by tuning the duration and the recovery time of the voltage pulse.

Naturally, the question arises about an approach allowing generating bias pulses of required strength and duration. The most plausible solution for this problem is the photoconducting switch,\cite{Auston-IEEE1983} employing the illumination of the semiconductor by a femtosecond laser pulse. The rise time of such switches is mostly defined by the rise time of the laser pulse, and the $RC$ characteristic of the circuit. The controllable and short decay time of the voltage pulse is, however, the challenging issue. We are aware of the reports where the voltage pulse durations down to several hundreds femtoseconds were achieved by designing special photoswitch circuits and using the semiconducting media characterized by short electron decay times.\cite{Dykaar-OQE1996,Wang-APL1995,Holzman-APL2001}

As for the decay time, they can be very small in metal-based structures. In the limit of ballistic transport estimates based on Eq.\,\ref{escape} give the escape time as small as $\tau_s\sim 10^{-14}$\,s for s-electrons in the point contact (Fig.\,\ref{fig2}(a)) of a size of $L \sim 10$\,nm. This value is the estimate of the upper limit of a sharpness of any equilibration process within the ballistic point contact. In particular, it describes the cooling rate after the external bias has been switched off. In the diffusive point contact (Fig.\,\ref{fig2}(a)) of a size $L\sim30$\,nm the escape time will be $\tau_s\sim10^{-12}$\,s for the electron mean free path of 3\,nm. For the tunnel structure presented on Fig.\ref{fig2}(b) one can control $\tau_s$ (Eq.\,\ref{escape-barrier}) by a proper choice of the tunneling transparency coefficient $k$. Certainly, to ensure effective control of the magnetization the time $\tau_s$ should not be smaller than the exchange scattering times $\tau_{e(A,B)}$. Thus, we emphasize that the advantage of the devices suggested is a possibility to control the process of switching in rather broad region by a proper choice of the device parameters including its geometry and the bias applied.

Therefore, we can conclude that the mechanism of formation of very short electric pulses seems to be the only restriction of operation times for the metal-based devices in question. However the same restriction concerns any electronic device while many other restrictions typical for semiconductor-based devices seem to be lifted.

An important ingredient of any spintronic device is a possibility of read-out of the magnetization controlled by some external stimulus. We believe that such a read-out in the case of the electric bias driven reversal can be provided by the well-known anomalous Hall-effect\cite{Nagaosa-RMP2010} (AHE). Indeed, the Hall voltage $V_H$ is related to the current $I$ through the structure as
$V_H/I=RH+ R_1M$, where $R$ and $R_1$ are some material-dependent parameters, $H$ is an external magnetic field , and $M$ is the sample magnetization.
In the absence of external field the magnitude and the sign of $V_H$ is controlled by the sample magnetization. Thus a presence of Hall
probes attached to the corresponding ferromagnetic layer allows to detect the state of the layer magnetization. Note that, although the AHE is often used to detect a presence of ferromagnetic ordering, when other technique possessed poor sensitivity for the
decisive conclusion. However, to the best of our knowledge it was not used to detect the sign of magnetization
since typically the samples had multidomain structure.

\section{Conclusions}

To conclude, we proposed the general microscopical model of the ultrafast magnetization reversal in antiferromagnetically coupled ferromagnetic metallic subsystems.  In the proposed model the rapid increase of the temperature of mobile $s-$ or $p-$electrons triggers effective exchange scattering between these electrons and the ferromagnetic subsystems of localized $f-$ and nearly localized $d-$electrons. Then incoherent
spin exchange between the two (nearly-)localized ferromagnetic subsystems is mediated by the mobile electrons. Owing to the different exchange relaxation times for two involved ferromagnetic subsystems, there is a moment after the excitation, when one of the subsystems is completely demagnetized, while another one still possess finite magnetization. It is this moment when the reversal of the "faster" sublattice takes place.  The model succeeds to explain most of the main features of the all-optical magnetization reversal in ferrimagnetic metallic single phase or multilayered structures, reported recently by several groups. An important argument in favor of the model is the fact that the switching was observed only for conducting structures inevitably containing mobile carriers. Since the effect of the external excitation in the considered here process is limited to a transient heating of the mobile electron system, we also analyze a possibility to trigger the magnetization reversal by application of the voltage bias. The relevant structures are metallic point contacts or tunneling structures with embedded ferrimagnetic metallic systems. It is shown that such devices allow to control switching with a great accuracy. We also suggest to use the anomalous Hall effect to register the switching thus playing a role of reading probes.

\section{Acknowledgements}
We thank Dr. A. V. Kimel for insightful discussions. This work has
been supported by the Program No.\,1 of the Board of the Russian
Academy of Sciences and the Russian Foundation for Basic Research under the grant No. 16-02-00064a. A. M. K. acknowledges the support from
the Russian Government under the grant No.\,14.B25.31.0025.

\end{document}